\documentclass[journal=jpclscd,manuscript=article]{achemso}
\usepackage[version=3]{mhchem} 
\usepackage{multirow}
\usepackage{hyperref}
\usepackage{array}
\usepackage[super]{nth}
\usepackage{bm}
\usepackage{mathtools}
\usepackage{amssymb}
\usepackage{booktabs}

\author{Sofia Apergi}
\affiliation{Materials Simulation and Modelling, Department of Applied Physics, Eindhoven University of Technology, P.O. Box 513, 5600 MB Eindhoven, The Netherlands}
\alsoaffiliation{Center for Computational Energy Research, Department of Applied Physics, Eindhoven University of Technology, P.O. Box 513, 5600 MB Eindhoven, The Netherlands}
\author{Geert Brocks}
\affiliation{Materials Simulation and Modelling, Department of Applied Physics, Eindhoven University of Technology, P.O. Box 513, 5600 MB Eindhoven, The Netherlands}
\alsoaffiliation{Center for Computational Energy Research, Department of Applied Physics, Eindhoven University of Technology, P.O. Box 513, 5600 MB Eindhoven, The Netherlands}
\alsoaffiliation{Computational Chemical Physics, Faculty of Science and Technology, and MESA+ Institute for Nanotechnology, University of Twente, P.O. Box 217, 7500 AE Enschede, The Netherlands}
\author{Shuxia Tao}
\affiliation{Materials Simulation and Modelling, Department of Applied Physics, Eindhoven University of Technology, P.O. Box 513, 5600 MB Eindhoven, The Netherlands}
\alsoaffiliation{Center for Computational Energy Research, Department of Applied Physics, Eindhoven University of Technology, P.O. Box 513, 5600 MB Eindhoven, The Netherlands}
\email{S.X.Tao@Tue.nl}

\title[An \textsf{achemso} demo]
  {Calculating the Circular Dichroism of Chiral Halide Perovskites: A Tight-Binding Approach}

\begin{document}

\begin{abstract}
    Chiral metal halide perovskites have emerged as promising optoelectronic materials for emission and detection of circular polarized visible light. Despite chirality being realized by adding chiral organic cations or ligands, the chiroptical activity originates from the metal halide framework. The mechanism is not well understood, as an overarching modeling framework is lacking. Capturing chirality requires going beyond electric dipole transitions, the common approximation in condensed matter calculations. We present a density functional theory (DFT) parameterized tight-binding (TB) model, which allows us to calculate optical properties including circular dichroism (CD) at low computational cost. Comparing Pb-based chiral perovskites with different organic cations and halide anions, we find that the structural helicity within the metal halide layers determines the size of the CD. Our results mark an important step in understanding the complex correlations of structural, electronic and optical properties of chiral perovskites, and provide a useful tool to predict new compounds with desired properties for novel optoelectronic applications.
\end{abstract}



Circularly polarized light (CPL) is a vital ingredient in many emerging or existing technologies, for instance, in quantum computing \cite{sherson2006quantum,langford2011efficient}, data storage and encryption \cite{chang2016discovery}, biomolecular sensing \cite{muller2009luminescent,hendry2010ultrasensitive}, imaging \cite{baba2002development,salomatina2009multimodal}, and asymmetric photochemical synthesis \cite{feringa1999absolute,carr2012lanthanide}. Chiral metal halide perovskites (called chiral perovskites from here on) have recently garnered extensive attention due to their potential in applications requiring the manipulation of CPL. Chiral perovskites are formed by mixing common perovskite precursors, such as lead or tin halides, with chiral organic cations. This mixing results in the formation of lower dimensional metal halide lattices, often two-dimensional (2D), separated by chiral organic spacers. Chirality transfer arising from interfacing chiral cations with metal-halide layers, breaks the mirror symmetry in those layers \cite{dang2021chiral,ma2021chiral}. The materials then exhibit optical absorption and luminescence  that is sensitive to the circular polarization of the light, in other words chiral dichroism (CD), making them ideal candidates for applications that involve CPL emission and detection  \cite{duim2021chiral,dong2019chiral,dang2021chiral,ma2021chiral,zhang2021multifunctional,jiang2022circular,rong2023interaction,liang2023metal}. 

With the increasing interest in chiral perovskites, the understanding of the mechanisms that govern their optical properties is lagging somewhat behind. Most of the attention so far has been given to the spin-dependent electronic properties of chiral perovskites, specifically to chirality-induced spin selectivity (CISS) \cite{yu2019advances}. A theoretical study emphasizes an interplay between spin-orbit coupling (SOC) and chirality to explain CISS, and expands this explanation to CD \cite{yu2020chirality}. It has also been argued that similar phenomena may be found in non-chiral perovskites if their symmetry gives rise to Rashba splitting \cite{sercel2020circular}. Sidestepping the detailed mechanisms behind CISS or CD, Ref. \cite{jana2021structural} demonstrates a correlation between the size of the chirality-induced spin-splitting and the size of the in-plane asymmetric distortion of the metal halide octahedra, using crystallographic and first-principles studies. 

The mechanisms behind specifically the chiroptical activity in halide perovskites are unclear. Whereas SOC plays a foundational role in explaining CISS, its role in CD is uncertain. Ultimately structure, lack of symmetry in particular, should govern these effects, but whether the electronic and optical properties are decided by the same structural parameters is open to question. Atomistic simulations are ideal for identifying the structural changes chiral molecules introduce in achiral perovskites. However, it is currently challenging to calculate optical activity in solid state materials due to the absence of an established computational framework \cite{berova2000circular}, similar to the frameworks that are available for chiral molecular systems\cite{fortino2023simulation}.

In this work, we present a computationally efficient method for studying the optical properties of chiral perovskites, using a DFT-parametrized tight-binding (TB) model. The TB model allows us to calculate band structures and single particle wave functions, including the effects of SOC, at a low computational cost. From there, optical response functions, such as dielectric functions and absorption coefficients, are calculated. The routinely used electric dipole approximation is not sufficient for calculating the main property of interest here, the CD. The latter requires including higher order terms, i.e., magnetic dipole/electric quadrupole. By comparing four chiral perovskites with varying structural distortions, we determine the structural parameters that control the features of the CD and discuss the trends when changing the chiral cations or halide anions.

\begin{figure}[htb]
\centering
    \includegraphics[width=\linewidth]{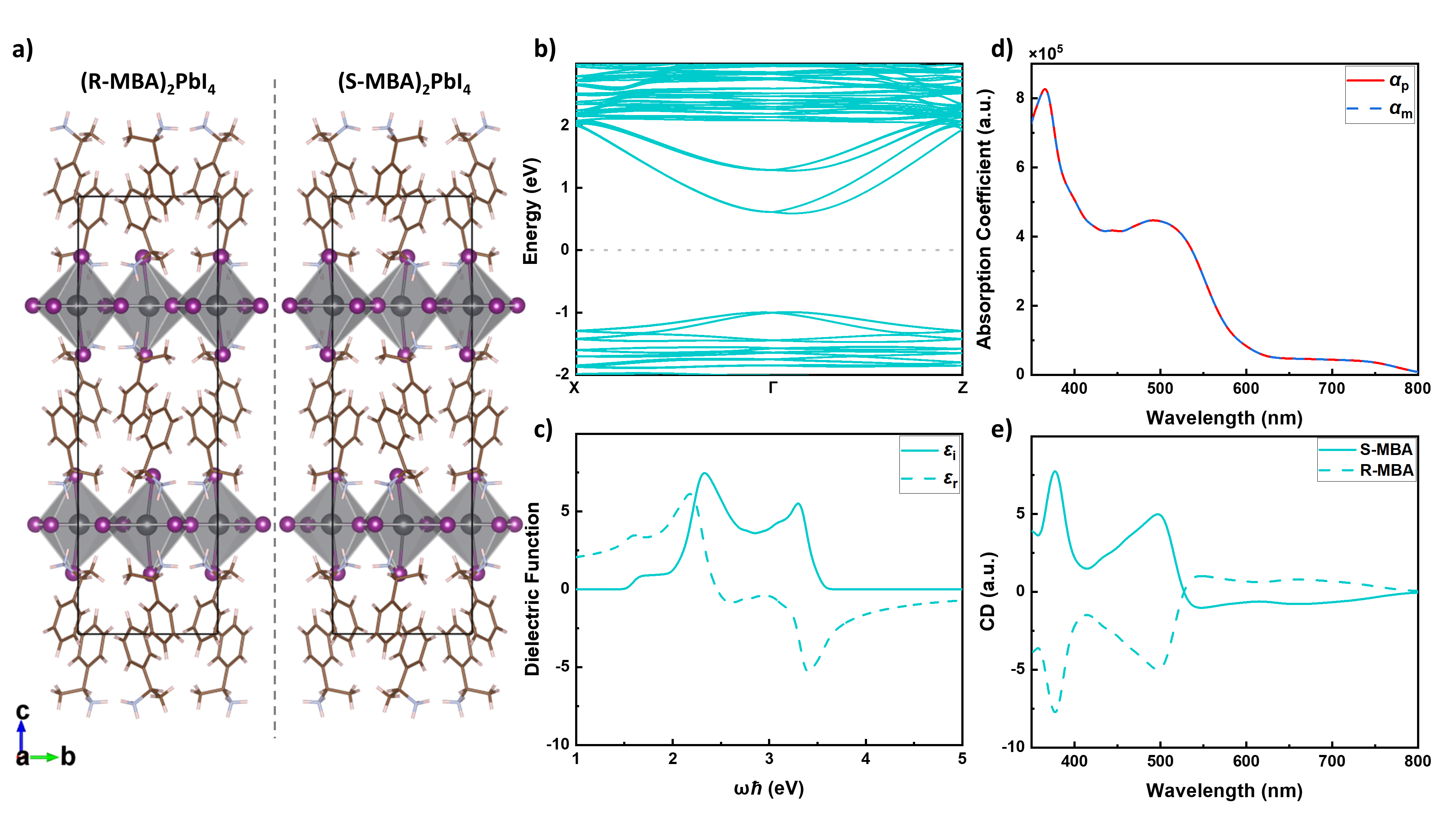}
    \caption{a) Atomistic representation of the R- (left) and S- (right) \ce{MBA2PbI4} perovskites. b) \ce{S-MBA2PbI4} band structure with SOC as calculated from TB. c) Real and imaginary dielectric function, d) absorption coefficient, and e) CD of the \ce{(R/S-MBA)2PbI4} perovskite calculated within the electric quadrupole - magnetic dipole approximation with the inclusion of SOC effects.}
    \label{fgr:1}
\end{figure}

We start from DFT calculations on archetype chiral 2D perovskites, \ce{(R/S-MBA)2PbI4}, with an orthorhombic unit cell consisting of four formula units. The structure is shown in Figure \ref{fgr:1}(a) as an example. It is fully relaxed using the PBE-D3-BJ exchange-correlation functional; details are given in the Computational Details section. The band structure calculated without SOC (Figure S1) is then used to construct a tight-binding (TB) model on the basis of Wannier orbitals, using the Wannier90 software \cite{mostofi2014updated}. The Wannier orbital basis of our TB model comprises Pb 6s and 6p, and I 5p orbitals. These are the orbitals that dominate the upper valence band and the lower conduction band states, see Figure S2. To account for SOC, on-site Hamiltonian matrices are added to the TB Hamiltonian matrix, which act on the Pb 6p and I 5p orbitals, respectively. Their general form is
\begin{equation}
    \mathcal{H}_{\mathrm{SOC}}^{\mathrm{B}} = \frac{\Delta_{\mathrm{SOC}}^{\mathrm{B}}}{3} \begin{pmatrix}
        0 & 0 & 0 & -1 & 0 & i \\
        0 & 0 & 1 & 0 & i & 0 \\
        0 & 1 & 0 & 0 & -i & 0 \\
        -1 & 0 & 0 & 0 & 0 & i \\
        0 & -i & i & 0 & 0 & 0 \\
        -i & 0 & 0 & -i & 0 & 0 \\
    \end{pmatrix},
    \label{eq:7.1}
\end{equation}
where $\mathcal{H}_{\mathrm{SOC}}^{\mathrm{B}}$ is presented on the basis of $\{\vert p_z \uparrow \rangle,\vert p_z \downarrow \rangle,\vert p_x \uparrow \rangle,\vert p_x \downarrow \rangle,\vert p_y \uparrow \rangle, \vert p_y \downarrow \rangle \}$. For B $=$ Pb, the SOC coupling parameter is $\Delta_{\mathrm{SOC}}^{\mathrm{Pb}} = 1.18$ eV, and for B $=$ I, it is $\Delta_{\mathrm{SOC}}^{\mathrm{I}} = 1.06$ eV, as proposed in Ref. \cite{cho2019optical}. In total, our TB model has a basis set consisting of 128 atomic orbitals.

The TB band structures are in very good agreement with the ones calculated with DFT; examples are given in Figure S2 (without SOC) and (with SOC) for \ce{(S-MBA)2PbI4}. The band gap calculated  without SOC (2.16 eV) is very close to the experimental value (2.14 eV) \cite{dang2020bulk}, but inclusion of SOC, in particular on Pb atoms, splits and shifts the lowest conduction band, and reduces the band gap considerably \cite{mosconi2013first}, in our case by 0.56 eV (Figure \ref{fgr:1}(b)). In principle, the experimental band gap can be recovered using hybrid exchange-correlation functionals or $GW$ calculations \cite{berger2018design}. However, the computational costs are high, so we refrain from this step. The effects of varying the band gap will be discussed later. The SOC-induced splitting of the conduction bands near $\Gamma$ stems from the breaking of mirror symmetry in chiral perovskites. Its effect on the optical properties will also be discussed below. 

Having constructed the TB model, we proceed with the calculation of the optical properties. In this paper, we focus on chiral dichroism (CD), which is basically the difference in the absorption between left- and right-handed CPL. The absorption coefficient is given by the expression
\begin{equation}
    \alpha_{\bm{q}\pm}(\omega) = \sqrt{2}\frac{\omega}{c} \sqrt{-\varepsilon_{r,{\bm{q}\pm}}(\omega) + \sqrt{\varepsilon_{r,{\bm{q}\pm}}^2(\omega)+\varepsilon_{i,{\bm{q}\pm}}^2(\omega)} }
    \label{eq:abs}
\end{equation}
where $\varepsilon_r$ and $\varepsilon_i$ are the real and imaginary parts of the relative electric permittivity; $\omega$, $\bm{q}$, and $\pm$ indicate the frequency, the wave vector and the polarization (right and left circular) of the electromagnetic radiation. We calculate the imaginary part $\varepsilon_i$ using the standard first-order perturbation theory independent particle expression, neglecting local field effects
\begin{equation}
    \varepsilon_{i,\bm{q}\pm} (\omega) = \frac{1}{4\pi \varepsilon_0} \left( \frac{2\pi e}{\omega}\right)^2 \sum_{\bm{k},\bm{k}'}^\mathrm{BZ} \vert\langle\mathrm{c}\bm{k}' \vert \bm{\hat{e}_\pm} \cdot \bm{v} \vert \mathrm{v} \bm{k} \rangle \vert^2 \delta(E_{\mathrm{c}}(\bm{k}') - E_{\mathrm{v}}(\bm{k}) - \hbar\omega).
    \label{eq:e_i}
\end{equation}
The real part $\varepsilon_r$ is then determined using a Kramers-Kroning transform. In \ref{eq:e_i} the indices c and v refer to wave functions of the conduction and valence band respectively, where $\bm{k},\bm{k}'$ refer to points in the Brillouin zone (BZ); $\bm{v}$ is the velocity operator of an electron (or hole), and $\bm{\hat{e}_\pm}$ is the polarization vector of the electromagnetic radiation; $E_{\mathrm{c}}(\bm{k}'),E_{\mathrm{v}}(\bm{k})$ are the electron and hole energies, respectively. The matrix elements $\langle\mathrm{c}\bm{k}' \vert \bm{\hat{e}_\pm} \cdot \bm{v} \vert \mathrm{v} \bm{k} \rangle$ are key quantities from a computational point of view. They are calculated in the approximation
\begin{multline}
    \langle\mathrm{c}\bm{k}' \vert \bm{\hat{e}_\pm} \cdot \bm{v} \vert \mathrm{v} \bm{k} \rangle \approx - \frac{i}{\hbar} \sum_{\alpha,\beta} c^*_{\mathrm{c}\beta}(\bm{k}) \left[\bm{\hat{e}_\pm} \cdot \nabla_{\bm{k}} \mathcal{H}_{0,\alpha\beta}(\bm{k})\right] c_{\mathrm{v}\alpha}(\bm{k}) \delta_{\bm{k}',\bm{k}} \\  - \frac{1}{\hbar} \sum_{\mathrm{n\neq c}} \frac{\left[ \sum_{\alpha,\beta} c^*_{\mathrm{c}\beta}(\bm{k}) \left[\bm{q} \cdot \nabla_{\bm{k}} \mathcal{H}_{0,\alpha\beta}(\bm{k})\right] c_{\mathrm{n}\alpha}(\bm{k}) \right] \left[\sum_{\alpha,\beta} c^*_{\mathrm{n}\beta}(\bm{k}) \left[\bm{\hat{e}}_\pm \cdot \nabla_{\bm{k}} \mathcal{H}_{0,\alpha\beta}(\bm{k})\right] c_{\mathrm{v}\alpha}(\bm{k}) \right]}{E_{\mathrm{c}}(\bm{k}) - E_{\mathrm{n}}(\bm{k})} \delta_{\bm{k}',\bm{k}}.
    \label{eq:eqmd} 
\end{multline}
Here $c_{\mathrm{n}\alpha}(\bm{k})$ and $E_{\mathrm{n}}(\bm{k})$  are the TB eigenvectors and eigenvalues respectively, with $\mathrm{n}$ the band index, and $\mathrm{v}$,$\mathrm{c}$ particular values of $\mathrm{n}$; $\mathcal{H}_{0,\alpha\beta}$ is the TB Hamiltonian matrix, whose gradient with respect to $\bm{k}$ can be determined analytically; indices $\alpha$ and $\beta$ refer to the orbitals that constitute the basis of the TB model. 16 bands are considered in the calculations, 8 valence and 8 conduction bands, and a $50 \times 50 \times 1$ k-point grid is used.

Typical calculations of optical properties only use the first term on the right-hand side of \ref{eq:eqmd}, which corresponds to the electric dipole (ED) approximation. Electric dipole transitions give by far the dominant contributions to optical absorption, but they are insufficient for calculating CD. The ED approximation assumes a homogeneous electric field across the sample, and such a field does not make a distinction between left and right rotation. Restoring the spatial wave character of the electromagnetic field to first order allows one to make this distinction, and can be used to calculate CD \cite{lan84}. This corresponds to the electric quadrupole/magnetic dipole (EQMD) approximation, which is represented by the second term on the right-hand side of \ref{eq:eqmd}. The derivation of \ref{eq:eqmd} can be found in the Supplementary Information.

Figure S3 illustrates that, while the ED approximation provides a fairly accurate dielectric function and absorption coefficient, the corresponding CD is zero. Including the EQMD term, the dielectric function and absorption spectrum are almost unchanged. That is because the EQMD matrix element is weaker than the ED matrix element by a factor of lattice vector/wavelength $\approx 10^{-3}$  \cite{peter2010fundamentals}. However, whereas the EQMD term is relatively unimportant in calculating the dielectric function and the absorption spectrum, it does supply the CD spectrum. Such a comparison is shown in Figure 3.

The calculated dielectric function of \ce{R/S-MBA2PbI4} is shown in Figure \ref{fgr:1}(c). Its imaginary part $\varepsilon_i(\omega)$ is reminiscent of the joint density of states (JDOS), with an onset at $\hbar\omega=1.6$ eV, and distinct peaks at $\sim 2.3$ eV and $3.3$ eV. The onset corresponds to the band gap as calculated with DFT. Van Hove singularities tend to be prominent in quasi-2D band structures, and markedly contribute to peaks in the JDOS \cite{peter2010fundamentals}. As \ce{R/S-MBA2PbI4} is indeed a quasi-2D structure, it suggests that the peaks at $\sim 2.3$ and $3.3$ eV correspond to  Van Hove singularities in the JDOS for transitions from the upper valence bands to the two lowest conduction bands.   

The absorption coefficient of \ce{R/S-MBA2PbI4}, calculated from \ref{eq:abs}, is shown in Figure \ref{fgr:1}(d). Qualitatively, the onset and peak structure of the absorption coefficient follow that of $\varepsilon_i(\omega)$, with an enlarged high frequency part, and the peaks shifted to a slightly higher frequency, because of the influence of $\varepsilon_r(\omega)$ in \ref{eq:abs}. The calculated CD spectrum, Figure \ref{fgr:1}(e), shows the typical oscillations also observed in experimentally measured CD spectra. In our case the extrema correspond to the peaks in the absorption spectrum. Figure \ref{fgr:1}(e) also demonstrates that the calculated CD signal of \ce{S-MBA2PbI4} is minus that of \ce{R-MBA2PbI4}, as it should. As the calculations on the two enantiomers have been totally independent of one another, this serves as a consistency check on the results of the calculations. 

While the order of magnitude of the calculated CD is in good agreement with experimental data, a more direct comparison with experiments is unfortunately hard to make. Our model for calculating CD assumes a single crystal. Experimental CD spectra depend on the thickness of the studied films and are also sensitive to other parameters, such as in-plane homogeneity \cite{ahn2017new}. Specifically, CD depends on the relative directions of the electromagnetic wave vector and the crystal axes, so in a multi-crystalline film where the grains have a variety of orientations, one observes an averaged signa
As mentioned above, as our calculation starts from a DFT band structure, the band gap is too small, and correspondingly, the onset energy of optical absorption (1.58 eV) is smaller than in experiment (2.14 eV) \cite{dang2020bulk}. We have tested the influence of the band gap on the CD spectra, using a scissors operation to increase the TB band gap to the experimental value. Results are shown in Figure S4. Apart from an obvious shift of the absorption onset to a smaller wave length, the CD spectrum appears slightly more compressed, which is largely the result of the frequency-dependent prefactors in \ref{eq:abs} and \ref{eq:e_i}. Other than these two differences, the CD spectra for different band gaps are very similar. As our main focus in this paper is on the relation between perovskite structure and CD spectrum, we leave the band gap problem to a future study.  

\begin{figure}[htb]
\centering
    \includegraphics[width=\linewidth]{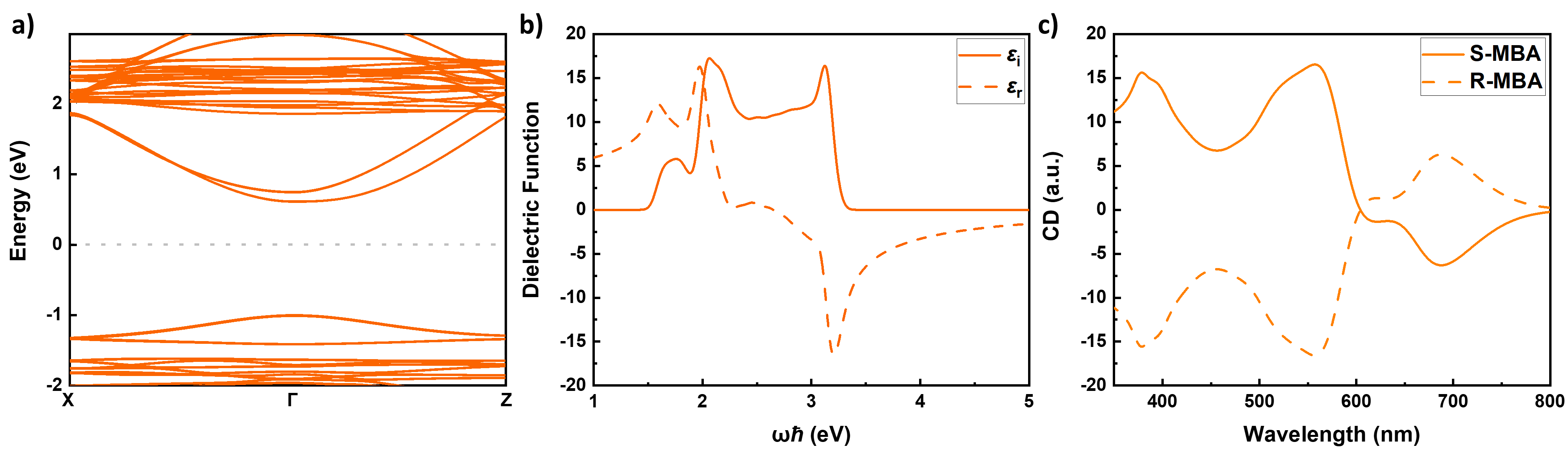}
    \caption{a) \ce{S-MBA2PbI4} band structure without SOC as calculated from TB. b) Real and imaginary dielectric function, and c) CD of the \ce{(R/S-MBA)2PbI4} perovskite calculated within the electric quadrupole - magnetic dipole approximation without the inclusion of SOC effects.}
    \label{fgr:2}
\end{figure}

SOC has played a central role in discussing CD in halide perovskites in the literature so far \cite{yu2020chirality,sercel2020circular,jana2021structural}. The results displayed in Figure \ref{fgr:1} are indeed calculated by including SOC. To examine how SOC affects the calculated CD, we repeat the calculation omitting the SOC matrices of \ref{eq:7.1} in the TB Hamiltonian. In this case, only 8 (4 valence and 4 conduction) bands are considered in the calculations, instead of 16 as was the case with SOC. The results of this calculation are shown in Figure \ref{fgr:2}. As is well-known, SOC  splits the bands, where the splitting is particularly large in the lowest conduction band, which has dominant Pb p character, Figure \ref{fgr:2}(a). Artificially enforcing the same band gap (1.58 eV) on the calculations with and without SOC (again, by means of a scissors operation), facilitates comparison of the optical spectra in Figure S5. In both cases, the structure of $\varepsilon_i(\omega)$ is mainly determined by that in the respective JDOSs, comparing Figures \ref{fgr:1}(c) and \ref{fgr:2}(b), and comparison is in Figure S4. The peaks observed here carry over to extrema in the CD, Figures \ref{fgr:1}(e) and \ref{fgr:2}(c). 

Omitting SOC still gives a non-zero CD, Figure \ref{fgr:2}(c), the size of which is at least of comparable magnitude to the CD calculated with SOC, Figure \ref{fgr:1}(e). We conclude from this that SOC is not vital to explain the occurrence of CD in halide perovskites, although it is important quantitatively. In fact, for \ce{R/S-MBA2PbI4} the maximum CD calculated without SOC is larger than that calculated with SOC. Apparently, in this case SOC dilutes the oscillator strengths of the optical transitions somewhat. 


Next, we look at whether there is a relation between the sizes of the CD and the SOC-induced spin splitting in the band structure. The latter determines the CISS effect, so it allows one to assess the correlation between CISS and CD. We calculate the CD of different perovskites with the same space group, but with spin splittings of different size. By doing so, differences in crystal symmetry  are excluded as the cause of variations in the optical properties. According to Ref. \cite{jana2021structural}, the chiral perovskites \ce{MHA2PbI4} and \ce{4-Cl-MBA2PbBr4} belong to the same space group as our benchmark \ce{MBA2PbI4}, but \ce{MHA2PbI4} has a much smaller spin-splitting, whereas \ce{4-Cl-MBA2PbBr4} has a much larger one. Besides having a different A cation, the latter compound also has a different halide anion, compared to the benchmark. To separate the effects of the two substitutions, we also construct an artificial \ce{4-Cl-MBA2PbI4} compound with the same crystal structure as \ce{4-Cl-MBA2PbBr4}, and reoptimize the cell parameters and ionic positions.

The optimized atomic structures of the studied chiral perovskites are shown in Figure S6. Since the three iodine-derived perovskites belong to the same space group and comprise the same inorganic layers, all structural differences stem from the different organic cations (Figure S7). Structural differences resulting from different anions then follow from substitution of iodine by bromine in \ce{4-Cl-MBA2PbX4}. Structural differences result in different electronic structures (in particular, the band splitting), as well as in a significant variation of the optical properties. The effect of SOC on the band structure of these compounds is typically characterized by $\Delta E_{max}$, which gives the maximum splitting in the lower conduction bands \cite{jana2021structural}. It ranges from $\Delta E_{max} = 0.04$ eV in \ce{MHA2PbI4} to 0.4 eV in \ce{4-Cl-MBA2PbI4}, with both \ce{MBA2PbI4} and \ce{4-Cl-MBA2PbBr4} having $\Delta E_{max} = 0.3$ eV, see Figure S7.

\begin{figure}[!b]
\centering
  \includegraphics[width=11cm]{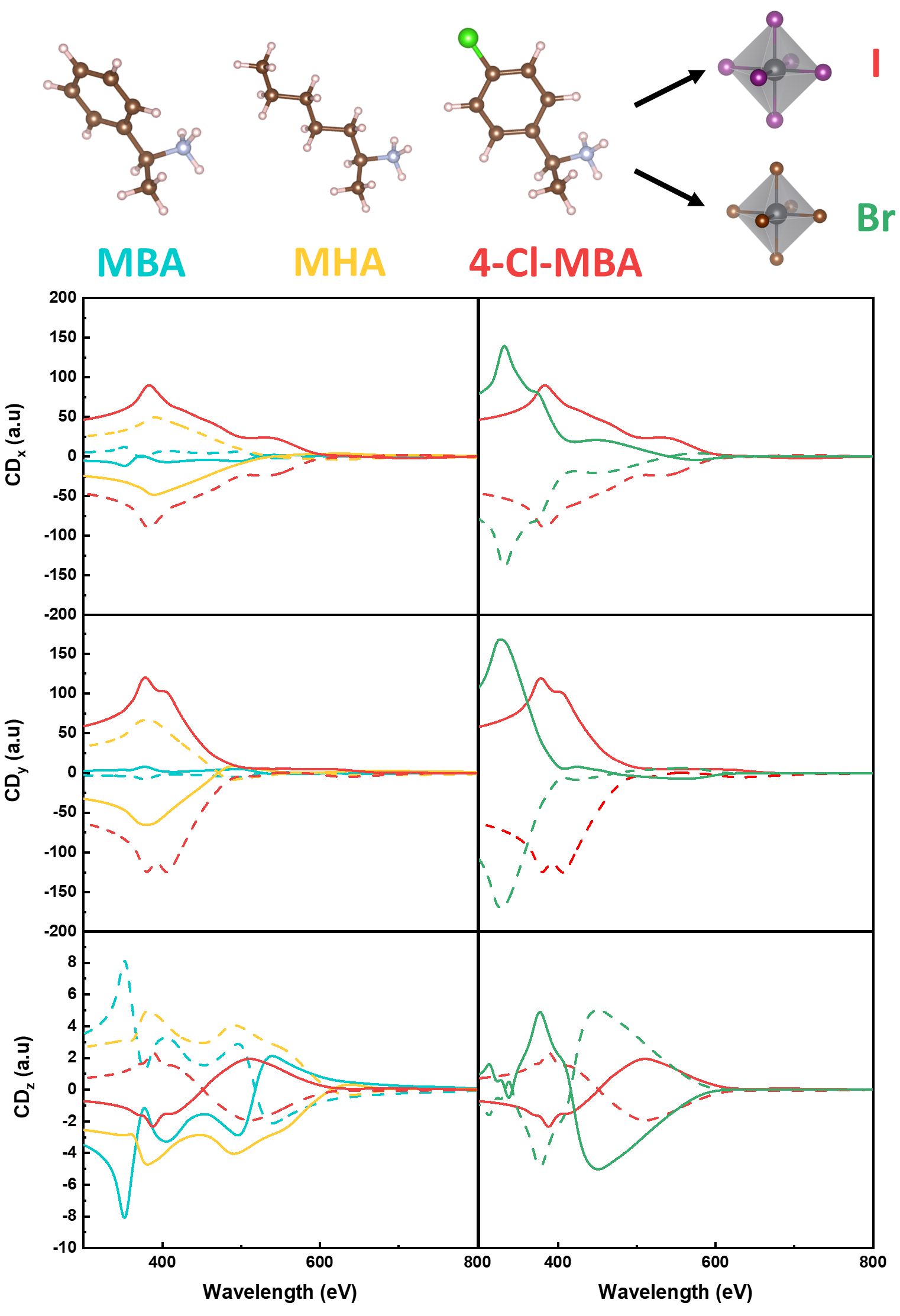}
  \caption{TB-calculated CD spectra for the chiral perovskites \ce{MHA2PbI4}, \ce{MBA2PbI4}, \ce{4-Cl-MBA2PbI4}, and \ce{4-Cl-MBA2PbBr4} including SOC, assuming that the light travels along the $x-$ (top), $y-$ (middle), and the $z-$ (bottom) direction. The calculated CD along the z-direction is much smaller than the x- and y-direction, but it is plotted in a different scale for clarity.}
  \label{fgr:3}
\end{figure}

Figure \ref{fgr:3} gives the CD spectra of the four compounds, \ce{MHA2PbI4}, \ce{MBA2PbI4}, \ce{4-Cl-MBA2PbI4}, and \ce{4-Cl-MBA2PbBr4}, for light traveling along one of the optical axes. The latter coincide with the crystal axes $\mathbf{a}$, $\mathbf{b}$ and $\mathbf{c}$ in an orthorhombic structure, denoted by cartesian directions $x$, $y$, $z$ in the following. Note that $x$ and $y$ are directions in the Pb-halide planes, and $z$ is perpendicular to those planes. The maximum spin-splitting discussed above occurs in the lower conduction bands dispersing in the $y$-direction \cite{jana2021structural}. 

According to Figure \ref{fgr:3}, of the iodide-based compounds \ce{4-Cl-MBA2PbI4} has the largest maximum CD in that direction, with that in \ce{MHA2PbI4} being a factor of two smaller, and the maximum CD of \ce{MBA2PbI4} more than ten times smaller than that of \ce{4-Cl-MBA2PbI4}. The maximum CD of the bromide-based compound \ce{4-Cl-MBA2PbBr4} is $\sim 50$ \% larger. There is no obvious correlation between the size of the CD, and that of the spin-splitting discussed above, which is another indication that spin effects are not driving the chiroptical activity.

Figure \ref{fgr:3} also shows the calculated CD spectra along the other two principal directions, $x$ and $z$. Interestingly, along the $x$-direction, the size of the CD is comparable to that in the $y$-direction, while it also follows the same trend n the order MBA $<$ MHA $<$ \ce{4-Cl-MBA}. In fact, the CD spectra of all compounds along the $x$ and $y$ directions are fairly similar. ]The main difference is that for all compounds the maximum CD is slightly lower along the $x$ than along the $y-$direction, apart from the CD of \ce{MBA2PbI4}, which is higher along $x$. 

Comparing different compounds, one observes that, apart from the amplitude, the CD spectra in $x$ and $y$ directions of the iodide-based ones are quite similar. In comparison, the spectrum of the bromide-based compound is shifted to a smaller wave length, as a result of the increased band gap of this compound. The CD spectra in the $z$ direction (perpendicular to the Pb halide planes) are markedly different from those in the in-plane directions. For \ce{MHA2PbI4}, \ce{4-Cl-MBA2PbI4}, and \ce{4-Cl-MBA2PbBr4}, the amplitude of the CD in the $z$ direction tends to be at least an order of magnitude smaller. Only for \ce{MBA2PbI4} the CD is comparable in size to that in the $x$ and $y$ directions, but for this compound the CD  in all directions are small anyway, as compared to the other compounds.

One would like to find a correlation between the structure of the compound and the amplitude of the CD. Inspired by Ref. \cite{jana2021structural}, seven different structural parameters, based on Pb halide bond angles and distances, are calculated, and the results are listed in Table S1. In agreement with Ref. \cite{jana2021structural}, we find that the spin splitting correlates with the parameter $\Delta \beta$, which is the difference betweentwo adjacent in-plane \ce{Pb-X-Pb} bond angles $\beta'$ and $\beta''$ (Figure S8). However, none of these structural parameters follow the same trend as the calculated CD. 

\begin{figure}[htb]
\centering
  \includegraphics[width=\textwidth]{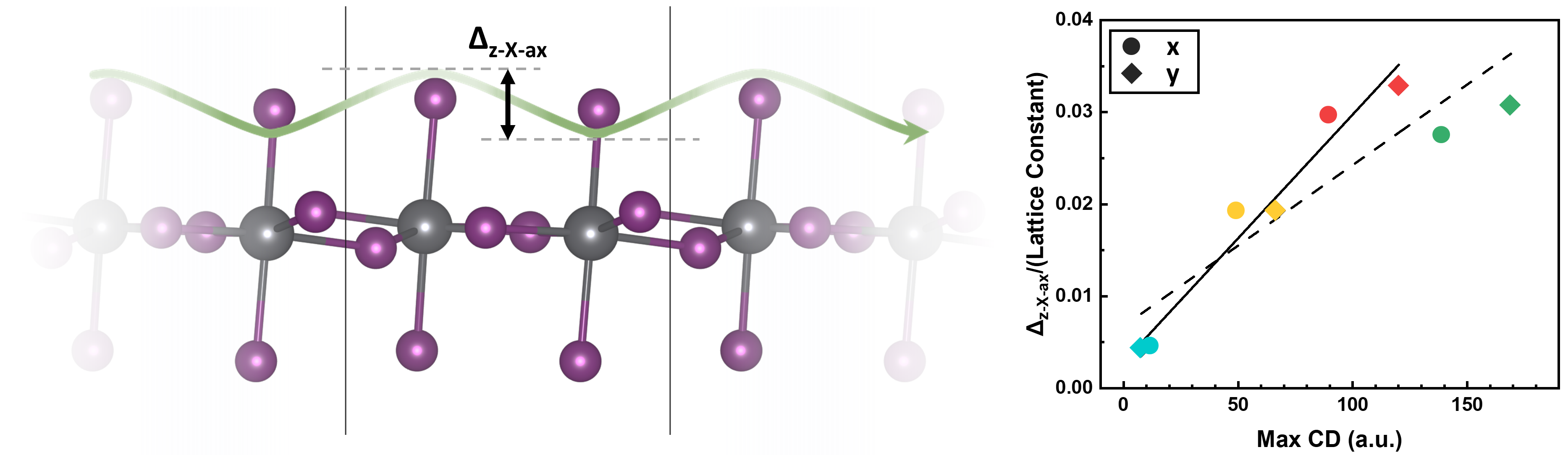}
  \caption{{PbI4} layer of \ce{4-Cl-MBA2PbI4} perovskite seen along the y-axis (left) and plot of the structural parameter $\Delta_{\mathrm{z-X-ax}}$ divided by the lattice parameter vs. maximum CD (right). The solid and dashed lines correspond to linear fitting without and with taking the CD of the Br-based perovskite into account.}
  \label{fgr:4}
\end{figure}

This implies that other features of the structures should be connected to the CD. Zooming in on the structure of a single layer of Pb halide octahedra, Figure \ref{fgr:4}, one observes that the equatorial halide ions do not exactly lie in one plane, nor do the axial halide ions. Based on this observation we define three simple structural parameters, $\Delta_{\mathrm{z-Pb-eq}}$, $\Delta_{\mathrm{z-X-eq}}$, and $\Delta_{\mathrm{z-X-ax}}$, which correspond to the amplitude of the displacements along the $z-$direction of the equatorial Pb, equatorial halide, and axial halide atoms, respectively (Figure \ref{fgr:4}). As can be seen in Table \ref{table:7.1}, while $\Delta_{\mathrm{z-Pb-eq}}$ does not really correlate with the CD, the other two structural parameters follow the same trend as the maximum CD along $x$ and $y$, for the three iodide-based perovskites. This means that they increase in the order of MBA $<$ MHA $<$ \ce{4-Cl-MBA}. 

According to the same table, smaller lattice parameters also tend to give rise to a larger CD. Interestingly, we can unify the CD amplitudes observed in the $x$- and $y$-directions. If we divide the structural parameter $\Delta_{\mathrm{z-X-ax}}$ by the lattice parameters $a$ and $b$ and plot it against the maximum CD along the $x$- and $y$- directions respectively, we get a trend that is very close to linear, see Figure \ref{fgr:4}. This construction does not work for the other two parameters, $\Delta_{\mathrm{z-Pb-eq}}$ and $\Delta_{\mathrm{z-X-eq}}$, see Figure S9. 

The maximum CD along the $z$-direction also tends to increase with decreasing lattice parameter. Apart from this simple observation, we have not been able to identify a structural parameter that correlates exactly with this trend. However, the CD amplitude in the $z$ direction is relatively small, compared to those in the $x$ and $y$ directions.

\begin{table}
\centering
\resizebox{\textwidth}{!}{
\begin{tabular}{cccccccccc}
\toprule
\textbf{Spacer} & CD$_{\mathrm{x,max}}$ & CD$_{\mathrm{y,max}}$ & CD$_{\mathrm{z,max}}$ & $\Delta_{\mathrm{z-Pb-eq}}$ &  $\Delta_{\mathrm{z-X-eq}}$ & $\Delta_{\mathrm{z-X-ax}}$ & $\bm{a}$ & $\bm{b}$ & $\bm{c}$ \\
\midrule
MBA & $11.93$ & $7.74$ & $8.10$ & $0.05$ & $0.03$ & $0.04$ & $8.74$ & $9.12$ & $28.60$ \\
MHA & $49.58$ & $66.71$ & $4.94$ & $0.17$ & $0.09$ & $0.17$ &  $8.81$ & $8.80$ & $33.72$ \\
4-Cl-MBA (I) & $89.90$ & $120.09$ & $2.33$ & $0.17$ & $0.52$ & $0.27$ &  $9.09$ & $8.22$ & $34.95$ \\
4-Cl-MBA (Br) & $139.43$ & $169.10$ & $5.02$ & $0.12$ & $0.52$ & $0.24$ &  $8.70$ & $7.79$ & $35.40$ \\
\bottomrule
\end{tabular}
}
\caption{CD maxima along the three principal directions, structural parameters $\Delta_{\mathrm{z-Pb-eq}}$, $\Delta_{\mathrm{z-X-eq}}$, and $\Delta_{\mathrm{z-X-ax}}$, and lattice parameters in \AA{} for the chiral perovskites \ce{MHA2PbI4}, \ce{MBA2PbI4}, \ce{4-Cl-MBA2PbI4}, and \ce{4-Cl-MBA2PbBr4}.}
\label{table:7.1}
\end{table}

Including \ce{4-Cl-MBA2PbBr4} one observes that, while the calculated CD obeys the same trend as the iodide-based compounds, described by $\Delta_{\mathrm{z-X-ax}}$ / lattice parameter. However, the relation is not exactly linear anymore (Figure \ref{fgr:4}). This can be partly explained by the fact that the increase in the band gap shifts CD peaks to smaller wave lengths and increases the size of the peaks slightly, a similar effect as is observed in Figure S5. This demonstrates that the features of the CD can not only be tuned by  changing the chiral ligands, but also by modifying the inorganic layers.

To summarize, in this work we present a method for calculating the optical properties of chiral halide perovskites. To ensure computational efficiency, our method is based on a TB model, parametrized from DFT calculations, and includes on-site SOC Hamiltonians. While similar approaches are used for the calculation of optical properties of various materials, most studies use the electric dipole approximation, which is not suitable for calculating chiral properties, such as CD. For this reason, we determine higher order terms, namely the electric quadrupole and magnetic dipole terms, which allow us to calculate the CDs of the chiral perovskites. 

We perform calculations on a number of chiral perovskites, varying either chiral ligands or halide anions, and analyse and compare their structural, electronic and optical properties. In contrast to frequent assumptions, we find no correlation between CD and the chirality-induced spin splitting, suggesting the two being controlled by different structural parameters. While in-plane inorganic metal-halide octahedral tilting distortions generally determine the size of the spin splitting, out-of-plane ionic displacements of metal-halide layer are the decisive factor for the size of CD. Specifically, the latter correlates with the out-of-plane displacement amplitude of the axial halide atoms, divided by the lattice parameter in the direction of the incident light. The chemical composition of the inorganic layer is another important factor in determining the CD. Replacing I with Br within the same structure increases the size of the CD, besides shifting the peaks in the CD spectrum. 

Our method is computational efficient and flexible, making it a suitable starting point for refinements to include effects that have not yet been considered, such as excitonic effects obtained from solving the Bethe-Salpeter equation, or the coupling with lattice dynamics to take the impact of finite temperatures into account. The availability of such a tool therefore creates opportunities to understand more complex interactions of light, spin and charge in this fascinating class of materials and to design new compositions with new functionalities.  

\section{Computational Details}
Density functional theory calculations were performed using the Projector Augmented Wave (PAW) method, as implemented in the Vienna Ab-Initio Simulation Package (VASP) \cite{Kresse1993,Blochl1994,Kresse1996,Kresse1999}. The electronic exchange-correlation interaction was described by the functional of Perdew, Burke, and Ernzerhof (PBE) within the generalized gradient approximation (GGA)\cite{perdew1996generalized} and energy and force convergence criteria of $10^{-5}$ eV and $0.02$ eV/{\AA} respectively were used in all calculations. The D3 correction with Becke-Jonson damping was employed to account for the van der Waals interactions due to the presence of the organic species \cite{grimme2010consistent}. The calculations were performed with a $4\times4\times1$ $\Gamma$-centered k-point grid and a kinetic energy cutoff of $500$ eV. Initial structures were acquired from Ref. \cite{jana2021structural} and they were subsequently optimized. During the geometry optimization, unit cells of four formula units were allowed to fully relax. In all cases the cell angles remained at 90$^{\circ}$, except for \ce{MBA2PbI4}, where after full relaxation, the cell angles slightly deviated from 90$^{\circ}$. In the latter case, the orthorhombic structure was reinforced, and the positions of the ions were allowed to relax.

\begin{acknowledgement}

S. Apergi acknowledges supports from funding NWO START-UP (Project No. 740.018.024) from the Netherlands.
S.T. acknowledges funding by NWO START-UP (Project No. 740.018.024) and VIDI (Project No. VI.Vidi.213.091) from the Netherlands. 

\end{acknowledgement}

\begin{suppinfo}

Chiral dichroism calculation, Figures S1-S8, Table S1

\end{suppinfo}

\bibliography{achemso-demo}

\end{document}